\documentclass{article}
\usepackage{spconf,graphicx}
\usepackage{nameref}
\usepackage[hidelinks]{hyperref}
\usepackage{soul}
\usepackage{color}
\usepackage{listings}
\usepackage[symbol]{footmisc}
\usepackage{dsfont}
\usepackage{amsmath}
\usepackage{mathtools}
\usepackage{centernot}
\usepackage{multirow}
\usepackage{makecell}
\usepackage{bm}
\usepackage{amsfonts}

\DeclarePairedDelimiter\abs{\lvert}{\rvert}%
\makeatletter
\let\oldabs\abs
\def\abs{\@ifstar{\oldabs}{\oldabs*}}

\newcommand{\norm}[1]{\left\lVert#1\right\rVert}

\newcommand{\parn}[1]{\left(#1\right)}

\newlength{\NOTskip}

\title{StyleTTS-VC: One-Shot Voice Conversion by Knowledge Transfer from Style-Based TTS Models}
\name{Yinghao Aaron Li, Cong Han, Nima Mesgarani}
\address{Department of Electrical Engineering, Columbia University, USA}

\begin{document}
\copyrightnotice{978-1-6654-7189-3/22/\$31.00~\copyright2023 IEEE}

%\ninept
%
\maketitle
\begin{abstract}
One-shot voice conversion (VC) aims to convert speech from any source speaker to an arbitrary target speaker with only a few seconds of reference speech from the target speaker. This relies heavily on disentangling the speaker's identity and speech content, a task that still remains challenging. Here, we propose a novel approach to learning disentangled speech representation by transfer learning from style-based text-to-speech (TTS) models. With cycle consistent and adversarial training, the style-based TTS models can perform transcription-guided one-shot VC with high fidelity and similarity. By learning an additional mel-spectrogram encoder through a teacher-student knowledge transfer and novel data augmentation scheme, our approach results in disentangled speech representation without needing the input text. The subjective evaluation shows that our approach can significantly outperform the previous state-of-the-art one-shot voice conversion models in both naturalness and similarity. 
\end{abstract}
\section{Introduction}
Voice conversion (VC) is a technique that converts one speaker’s voice into another’s voice while preserving linguistic and prosodic information such as phonemes and prosody. Recent advances in deep learning have enriched research on one particular type of voice conversion: one-shot voice conversion. This type of voice conversion, also known as any-to-any voice conversion, aims to convert speech from any source speaker to an arbitrary target speaker using only a few seconds of reference audio from the target speaker. To convert an unseen speaker’s voice into another speaker’s voice unseen during training, the model needs to learn a shared representation of speech across all potential sources and target speakers \cite{qian2019autovc}. Therefore, learning disentangled representations of speech and speaker identity is crucial for successful one-shot voice conversion. 

Several techniques have been proposed for learning disentangled representations, including instance normalization \cite{chou2019one, wu2020one, chen2021again},  vector quantization \cite{wu2020one, van2020vector, wang2021vqmivc, tang2022avqvc}, transfer learning from ASR or TTS models \cite{li2021ppg, lin2021s2vc, zhang2021transfer, casanova2022yourtts, gabrys2022voice}, and adversarial training \cite{wang2020one, tang2021tgavc}. These methods, albeit effective, do not guarantee that the empirically trained representations contain no source speaker information. VC systems such as Mellotron \cite{valle2020mellotron} and Cotatron \cite{park2020cotatron}, on the other hand, use phoneme alignment and pitch curve from the source speech and re-synthesize the speech of the target speaker. Since phoneme alignment and normalized pitch curve are largely speaker-agnostic, the re-synthesized speech should only reflect the speech content and prosody of the source audio without leaking any other source-specific information. These TTS-based methods that theoretically guarantee a disentangled representation still suffer from two essential problems. The major drawback of TTS-based models is that this method requires input text or a sequence of phonemes to generate the alignment which limits its potential for applications in real-time inference. Zhang et. al. \cite{zhang2021transfer} has made an attempt to address this problem by training an additional mel-spectrogram encoder that produces the same latent representation as the one generated from phoneme alignment and text representation. This is equivalent to training an automatic speech recognition (ASR) system, but as we show here, this way of encoder training is not optimal. Another obstacle endured by the TTS method is the generalization problem. Since the original TTS models are trained to only reconstruct speech from the pitch and phoneme alignment of the source speaker, there is no guarantee that the synthesized speech will sound natural and similar to the target speakers when the input pitch and phoneme alignment are from a different speaker.

\begin{figure*}[!th]
  \centering
  \includegraphics[width=\linewidth]{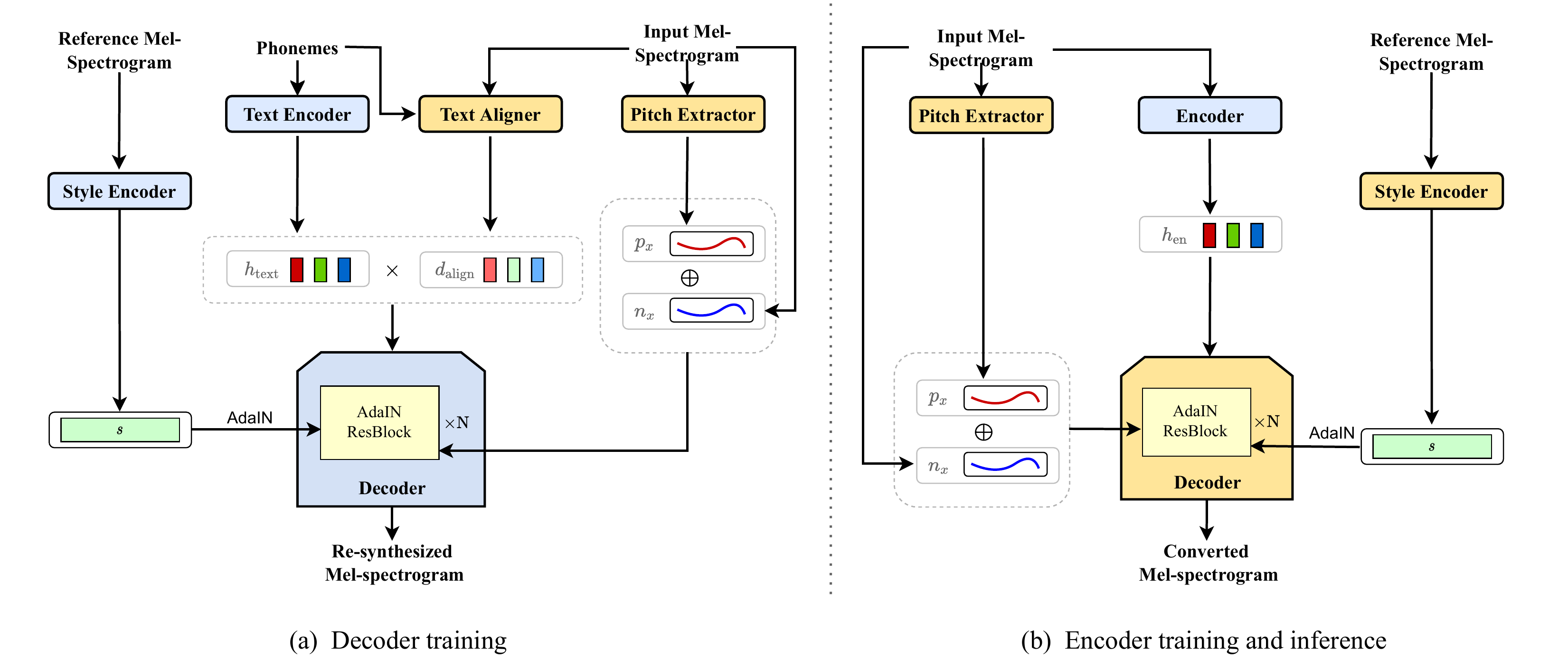}
  \caption{Training and inference schemes for StyleTTS-VC. The modules in blue are trained while those in orange are pretrained and hence fixed during training. (a) Step 1 of training where the decoder is trained to synthesize target speech from a reference mel-spectrogram and pitch curve, energy, phoneme alignment and text from an input mel-spectrogram. (b) Step 2 of training and inference procedures where the text aligner and text encoder are replaced by a mel-spectrogram encoder. }
  \label{fig:fig1}
\end{figure*}

In this paper, we present StyleTTS-VC, a non-parallel one-shot voice conversion framework based on StyleTTS \cite{li2022styletts}, a style-based text-to-speech model. We address the aforementioned generalization problems by first training a StyleTTS speech decoder with a cycle consistency loss function and adversarial objectives. We then train a mel-spectrogram encoder to produce representations that reconstruct the decoder output generated using representations from phoneme alignment for all speakers in the training set with both synthesized and real speech as input. Unlike the previous method \cite{zhang2021transfer}, our proposed technique does not force the encoded representations to be close to the phoneme alignment representations. The subjective human evaluation shows that our model outperforms the previous state-of-the-art one-shot voice conversion model, YourTTS \cite{casanova2022yourtts}, and two other baseline models, AGAIN-VC \cite{chen2021again} and VQMIVC \cite{wang2021vqmivc}, for unseen source and target speakers. Moreover, since our model consists of only convolutional layers without non-causal RNN or transformers, our model has the capability to perform real-time inference with a faster-than-real-time vocoder. 

Our work makes multiple contributions: (i) we show that the cycle consistency and adversarial objective are effective in training both TTS decoder and mel-spectrogram encoder for VC applications, (ii) we introduce novel data augmentation using text-guided voice conversion results as both input and target during training, and (iii) we demonstrate that the loss function proposed in \cite{zhang2021transfer} is suboptimal for transfer learning from TTS models for voice conversion applications and propose an alternative solution with a mutual information (MI) maximization objective. The audio samples from our model are available at \url{https://styletts-vc.github.io}.

\section{Methods}

\subsection{StyleTTS}
StyleTTS \cite{li2022styletts} is a style-based non-autoregressive text-to-speech model that integrates style information through adaptive instance normalization (AdaIN) \cite{huang2017arbitrary}. The StyleTTS framework consists of eight modules: text encoder, style encoder, discriminator, text aligner, pitch extractor, speech decoder, duration predictor, and prosody predictor. Since we only use the speech decoder for voice conversion, we only describe the modules and objectives needed to train the speech decoder here. We do not use the duration and prosody predictors because they are not relevant to the VC application. For simplicity, we assume that the text aligner and pitch extractor are pre-trained and fixed during training. An overview of StyleTTS decoder training is given in Figure \ref{fig:fig1}a. 

\noindent\textbf{Text encoder. } Given input phonemes $\bm{t}$, our text encoder $T$ encodes $\bm{t}$ into latent representation $\bm{h}_\text{text} = T(\bm{t})$. We use the same text encoder as in Tacotron  2 \cite{shen2018natural}.

\noindent\textbf{Style encoder. } Given an input mel-spectrogram $\bm{x}$, the encoder extracts the style code $s = S(\bm{x})$. For the voice conversion application, $s$ is roughly equivalent to the speaker embedding. The style encoder is the same as in StarGANv2-VC \cite{Li2021StarGANv2VCAD} without the domain-specific linear projection layers.

\noindent\textbf{Decoder. } The decoder $G$ synthesizes the mel-spectrogram $\hat{\bm{x}} = G\parn{\bm{h}_\text{text} \cdot \bm{d}_\text{align}, s, p_{\bm{x}}, n_{\bm{x}}}$ from an input audio $\bm{x}$, where $\bm{h}_\text{text} \cdot \bm{d}_\text{align}$ is the aligned latent representation of phonemes, $s$ is the style code of target speaker, $p_{\bm{x}}$ is pitch contour and $n_{\bm{x}}$ is the log norm (energy) of $\bm{x}$ per frame. Our decoder consists of seven residual blocks with AdaIN (equation \ref{eq:adain}), with which the style code $s$ is introduced into $G$. The $p_{\bm{x}}$ and $n_{\bm{x}}$ are normalized and concatenated with the output from every residual block as the input to the next residual block.

\begin{equation}
    \label{eq:adain}
    \text{AdaIN}(x, s) = L_\sigma(s) \frac{x - \mu(x)}{\sigma(x)} + L_\mu(s),
\end{equation} 
where $x$ is a single channel of the feature maps, $s$ is the style vector, $\mu(\cdot)$ and $\sigma(\cdot)$ denotes the channel mean and standard deviation, and $L_\sigma$ and $L_\mu$ are learned linear projections for computing the adaptive gain and bias using the style vector $s$. 

\noindent\textbf{Discriminator. } We employ the same discriminator $D$ as in StarGANv2-VC \cite{Li2021StarGANv2VCAD} for seen speakers during training. The discriminator has the same architecture as the style encoder but with the domain-specific linear projection layer for each speaker. The domain-specific layer helps the discriminator to capture detailed features of each speaker in the training set. 

\noindent\textbf{Text aligner and pitch extractor. } The text aligner $A$ is based on the decoder of Tacotron  2 with attention. It is pre-trained for automatic speech recognition (ASR) task on the LibriTTS corpus \cite{zen2019libritts}. The pitch extractor $F$ is a pre-trained JDC network \cite{kum2019joint} trained on LibriTTS with ground truth F0 estimated using Harvest \cite{morise2017harvest}. The text aligner is the same as the ASR model in \cite{Li2021StarGANv2VCAD}, and the pitch extractor is the same as the F0 network used in \cite{Li2021StarGANv2VCAD}. Both models are pre-trained and fixed during training. 

\subsection{StyleTTS-VC}
For voice conversion without text input, we train an additional encoder $E$ that encodes a mel-spectrogram $\bm{x}$ into $\bm{h}_\text{en}$ such that $ G\parn{\bm{h}_\text{text} \cdot \bm{d}_\text{align}, s, p_{\bm{x}}, n_{\bm{x}}} = G\parn{E(\bm{x}), s, p_{\bm{x}}, n_{\bm{x}}}$. That is, the encoder learns to produce representations that can be used by the decoder to synthesize the same speech as from the representations generated by the text encoder and phoneme alignment. 
The encoder consists of six 1-D residual blocks with instance normalization \cite{ulyanov2016instance}, similar to those used in \cite{chen2021again} and \cite{Li2021StarGANv2VCAD}. Unlike \cite{zhang2021transfer}, we do not enforce $\bm{h}_\text{en} = \bm{h}_\text{text} \cdot \bm{d}_\text{align}$, in which case the encoder becomes an ASR model and may produce unnatural speech. The effect of enforcing $\bm{h}_\text{en} = \bm{h}_\text{text} \cdot \bm{d}_\text{align}$ is examined in section \ref{section3.4}.

During inference, for any given input $\bm{x}_\text{in}$, we extract the pitch $p_\text{in} = F(\bm{x}_\text{in})$ and energy $n_{\text{in}_t} = \log \sqrt{\sum\limits_{n=1}^{N} \parn{{\bm{x}_{n, t}}}^2}$ where $\bm{x}_{n, t}$ represents the $n^{\text{th}}$ mel of the $t^{\text{th}}$ frame, $N$ the number of mels, and the speech content $\bm{h}_\text{en} = E(\bm{x}_\text{in})$. We compute the style code $s = S(\bm{x}_\text{ref})$ to synthesize $\hat{\bm{x}}_\mathrm{trg}$ from the target speaker. Since both $p_\text{in}$ and $n_\text{in}$ are 1-dimensional normalized curves, they cannot contain more information than pitch and volume. Since $\bm{h}_\text{en}$ is trained to replicate the effects of $\bm{h}_\text{text} \cdot \bm{d}_\text{align}$ for all possible speech generated by $G$, $\bm{h}_\text{en}$ is also a disentangled representation for phonemes that contain no speaker information. An overview of StyleTTS-VC is provided in Figure \ref{fig:fig1}b. 

\subsection{Training Objectives}
We train our model in two steps. We first train the decoder with a cycle consistency loss function, and we then train the encoder with the aforementioned objective with a fixed pre-trained decoder. Given a mel-spectrogram $\bm{x}\in \mathcal{X}_{y_\mathrm{src}}$, a reference $\bm{x}_\text{ref}\in \mathcal{X}_{y_\mathrm{trg}}$, the source speaker $y_\mathrm{src} \in \mathcal{Y}$ and the target speaker $y_\mathrm{trg} \in \mathcal{Y}$, we train our model with the following loss functions. 

\noindent\textbf{Mel reconstruction loss. } Given a mel-spectrogram $\bm{x} \in \mathcal{X}$ and its corresponding text $\bm{t} \in \mathcal{T}$, the decoder is trained with 
\begin{equation} \label{eq2}
\mathcal{L}_{rec} = \mathbb{E}_{\bm{x}, \bm{t}}\left[{\norm{\bm{x} - G\parn{\bm{h}_\text{text} \cdot \bm{d}_\text{align}, s, p_{\bm{x}}, n_{\bm{x}}}}_1}\right],
\end{equation}
where $\bm{h}_\text{text} = T(\bm{t})$ is the encoded phoneme representation, $\bm{d}_\text{align}$ is the attention alignment pre-computed from the text aligner, $s = S(\bm{x})$ is the style code of $\bm{x}$, $p_{\bm{x}} = F(\bm{x})$ is the pitch F0 of $\bm{x}$ and $n_{\bm{x}}$ is the energy of $\bm{x}$. We use the monotonic version of the attention alignment obtained by a dynamic programming algorithm \cite{kim2020glow} for 50\% of the time because the attention alignments are not strictly monotonic and can contain speaker information. 
% When training the decoder, we use the monotonic version of $\bm{d}_\text{align}$ through monotonic alignment search proposed in Glow-TTS \cite{kim2020glow} because the attention alignments is not strictly monotonic and may leak source speaker information. The alignments are pre-computed as our text aligner is fixed. 

\noindent\textbf{Style reconstruction loss. } To learn meaningful style code that represents speaker embeddings, we used a self-supervised style reconstruction similar to \cite{Li2021StarGANv2VCAD}
\begin{equation} \label{eq2}
\mathcal{L}_{sty} = \mathbb{E}_{ \bm{x}, \bm{t}, {\bm{x}_\text{ref}}}\left[{\norm{S({\bm{x}_\text{ref}}) - S(\hat{\bm{x}}_\mathrm{trg})}_1}\right],
\end{equation}
where $\hat{\bm{x}}_\mathrm{trg} = G\parn{\bm{h}_\text{text} \cdot \bm{d}_\text{align}, S(\bm{x}_\text{ref}), p_{\bm{x}}, n_{\bm{x}}}$, the reconstructed mel-spectrogram under style code of ${\bm{x}_\text{ref}}$ with the phonemes, alignment, pitch, and energy information of $\bm{x}$. 

\noindent\textbf{Encoder loss. } When training the encoder, we require the encoder to produce representations that can be used by the decoder to produce the same speech as those generated using representations through phoneme alignment under the encoder loss for an arbitrary target speaker in the training set
\begin{equation} \label{eq:4}
\mathcal{L}_{en} = \mathbb{E}_{\bm{x}, \bm{t}, \bm{x}_\text{ref}}\left[{\norm{\hat{\bm{x}} - G\parn{E(\bm{x}), S(\bm{x}_\text{ref}), p_{\bm{x}}, n_{\bm{x}}}}_1}\right],
\end{equation}
where $\hat{x} = G\parn{\bm{h}_\text{text} \cdot \bm{d}_\text{align}, S(\bm{x}_\text{ref}), p_{\bm{x}}, n_{\bm{x}}}$ the converted speech using text representation and phoneme alignment. 

Here $\bm{x}$ can be either ground truth from the training set or synthesized data. $\bm{x} = G\parn{\hat{\bm{h}}_\text{text} \cdot \hat{\bm{d}}_\text{align}, S(\hat{\bm{x}}_\text{ref}), p_{\hat{\bm{x}}}, n_{\hat{\bm{x}}}}$ when $\bm{x}$ is synthesized, where $\hat{\bm{h}}_\text{text}$ and $\hat{\bm{d}}_\text{align}$ are text and alignment of another speech sample $\hat{\bm{x}}$ and $\hat{\bm{x}}_{\text{ref}}$ is another reference audio different from $\bm{x}_{\text{ref}}$. That is, when $\bm{x}$ is synthesized, it is a converted speech sample used as an input. This novel data augmentation fully explores the input and target space of the pre-trained TTS decoder and produces more robust models compared to those trained without this technique. 

\noindent\textbf{Phoneme loss. } Since we do not demand $E(\bm{x}) = \bm{h}_\text{text} \cdot \bm{d}_\text{align}$, there is no guarantee that the generated speech keeps the original phoneme content. We employ a phoneme loss function to maximize the mutual information (MI) \cite{boudiaf2020unifying} between the encoded representations and the phonetic content through a linear projection $P$ for each frame of the input

\begin{equation} \label{eq3}
\mathcal{L}_{MI} = \mathbb{E}_{\bm{x}, \bm{t}}\left[\frac{1}{T}\sum\limits_{i=1}^T{\textbf{CE}\parn{\parn{\bm{d}_\text{align} \cdot \bm{t}}_i, \parn{P \cdot E(\bm{x})}_i}}\right],
\end{equation}
where $T$ is the number of frames and $\textbf{CE}(\cdot)$ denotes the cross-entropy loss function.

\noindent\textbf{Cycle consistency loss. } To make sure that the decoder generalizes to different input style codes independent of text, pitch and energy, we also employ a cycle consistency loss function 
\begin{equation} \label{eq2}
\mathcal{L}_{cycle} = \mathbb{E}_{\bm{x}, \bm{t}}\left[{\norm{\bm{x} - G\parn{\bm{h}, S(\bm{x}), p_{{\hat{\bm{x}}_\mathrm{trg}}}, n_{{\hat{\bm{x}}_\mathrm{trg}}}}}_1}\right],
\end{equation}
where $p_{{\hat{\bm{x}}_\mathrm{trg}}}$ is the pitch curve and $n_{{\hat{\bm{x}}_\mathrm{trg}}}$ is the energy of the converted speech $\hat{\bm{x}}_\mathrm{trg}$. When training the decoder, $\bm{h} = \bm{h}_\text{text} \cdot \hat{\bm{d}}_\text{align}$ and $\bm{x}$ is the ground truth where $\hat{\bm{d}}_\text{align}$ is the attention alignment of the converted speech $\hat{\bm{x}}_\mathrm{trg}$. When training the encoder, $\bm{h} = E(\bm{x})$ and $\bm{x}=\hat{\bm{x}}$ in equation \ref{eq:4}.

\noindent\textbf{Adversarial loss.} We use two adversarial objectives: the original cross-entropy loss function for adversarial training following \cite{Li2021StarGANv2VCAD} and the additional feature-matching loss following \cite{kong2020hifi}

\begin{equation}
\begin{aligned}    
  \mathcal{L}_{adv} =\text{ } &\mathbb{E}_{\bm{x}, y_\mathrm{src}}\left[\log D(\bm{x}, y_\mathrm{src})\right] +\\
  &\mathbb{E}_{\bm{x}, \bm{t}, y_\mathrm{trg}}\left[\log\parn{1 - D(\hat{\bm{x}}_\mathrm{trg}, y_\mathrm{trg})}\right],
\end{aligned}
  \label{eq0}
\end{equation}

\begin{equation} \label{eq5}
\mathcal{L}_{fm} = \mathbb{E}_{\bm{x}, \hat{\bm{x}}}\left[\sum\limits_{l=1}^{L} \frac{1}{N_l}\norm{D^l(\bm{x}, y_\mathrm{src}) - D^l(\hat{\bm{x}}, y_\mathrm{src})}_1\right],
\end{equation}
where $D(\cdot, y)$ denotes the output of discriminator for the speaker $y \in \mathcal{Y}$, $\hat{\bm{x}}_\mathrm{trg} = G\parn{\bm{h}, S(\bm{x}_\text{ref}), p_{\bm{x}}, n_{\bm{x}}}$ the converted speech, $\hat{\bm{x}} = G\parn{\bm{h}, s, p_{\bm{x}}, n_{\bm{x}}}$ the reconstructed speech, $L$ is the total number of layers in $D$ and $D^l$ denotes the output feature map of $l^\text{th}$ layer with $N_l$ features. The values of $\bm{x}$ and $\bm{h}$ are the same as in equation \ref{eq2} depending on whether the encoder $E$ or the decoder $D$ is trained.

\noindent\textbf{Full objectives. } Our full  objective functions for training the decoder can be summarized as follows: 
\begin{equation} \label{eq6}
\begin{aligned}  
\min_{G, T, S} \max_{D} \text{    }
& \mathcal{L}_{rec} + \lambda_{sty}\mathcal{L}_{sty} + \lambda_{cycle}\mathcal{L}_{cycle}  \\
& + \lambda_{adv}\mathcal{L}_{adv} + 
\lambda_{fm}\mathcal{L}_{fm},
\end{aligned}
\end{equation}
and full objective functions for the encoder are: 
\begin{equation} \label{eq7}
\begin{aligned}  
\min_{E, P} \max_{D} \text{    }
& \mathcal{L}_{en} + \lambda_{cycle}\mathcal{L}_{cycle} + \lambda_{MI}\mathcal{L}_{MI}  \\
& + \lambda_{adv}\mathcal{L}_{adv} + 
\lambda_{fm}\mathcal{L}_{fm}.
\end{aligned}
\end{equation}
% where $\lambda_{rec}, \lambda_{sty}, \lambda_{cycle}, \lambda_{adv}, \lambda_{fm}, \lambda_{en}$ and $\lambda_{asr}$ are hyperparameters for each term

\section{Experiments}

\begin{figure*}[t]
  \centering
  \includegraphics[width=\linewidth]{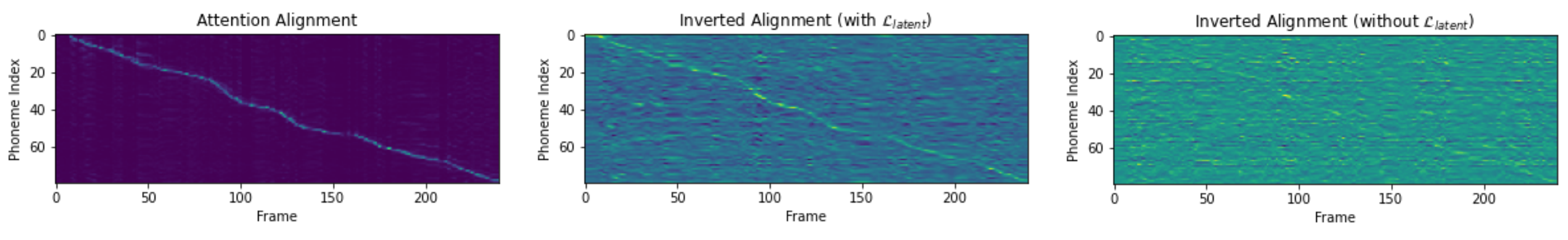}
  \caption{Example of attention alignment $d_\text{align}$ and inverted alignments obtained through $E(\bm{x}) \cdot \bm{h}_\text{text}^{-1}$ where $\bm{h}_\text{text}^{-1}$ is a pseudoinverse of $\bm{h}_\text{text}$. 
  The representation trained with $\mathcal{L}_{latent}$ clearly reproduces the monotonic alignment, indicating that $E$ acts like an ASR model. % On the other hand, the encoder trained without $\mathcal{L}_{latent}$ does not reproduce the monotonic alignment. 
  }
  \label{fig:fig2}
\end{figure*}

\subsection{Datasets} We used the VCTK \cite{yamagishi2019cstr} corpus to evaluate our models. The VCTK dataset consists of 109 native English speakers with various accents, each of which reads approximately 400 sentences. We followed the same procedure described in \cite{wang2021vqmivc}, where the 89 speakers were randomly selected for training and the rest 20 speakers were used as unseen speakers for testing. We further divided samples of the selected 89 speakers into training and validation sets with a 90\%/10\% split. The samples were downsampled to 24 kHz. 
We converted the text sequences into phoneme sequences using an open-source tool \footnote{\url{https://github.com/Kyubyong/g2p}}. We extracted mel-spectrograms with a FFT size of 2048, hop size of 300, and window length of 1200 in 80 mel bins using TorchAudio \cite{yang2021torchaudio}.
The generated mel-spectrogram was converted into waveforms using the Hifi-GAN \cite{kong2020hifi} and downsampled to 16 kHz to match the baseline models.

\subsection{Training Details} We first trained the decoder for 100 epochs with $\lambda_{sty} = 0.2$, $\lambda_{cycle} = 1$, $\lambda_{adv} = 1$ and $\lambda_{fm} = 0.2$, and we then trained the encoder for 100 epochs with $\lambda_{MI} = 1$. We trained both models using the AdamW optimizer \cite{loshchilov2018fixing} with $\beta_1 = 0, \beta_2 = 0.99$, weight decay $\lambda = 10^{-4}$, learning rate $\gamma = 10^{-4}$ and batch size of 64 samples. We randomly divided mel-spectrograms into segments of the shortest length in the batch.

\begin{table}[!t]
	\centering
	\caption{Comparison of MOS with 95\% confidence intervals between different models. \\ } 
    \begin{tabular}{c|c|c}
    \hline
    Method & MOS-N & MOS-P \\
    \hline
    Ground Truth      & 4.68 ($\pm$ 0.05)  & 4.58 ($\pm$ 0.07) \\
    StyleTTS-VC   & \textbf{3.89} ($\pm$ \textbf{0.09})  &    \textbf{3.66} ($\pm$ \textbf{0.10}) \\
    YourTTS & 3.70 ($\pm$ 0.10) & 3.45 ($\pm$ 0.10) \\
    VQMIVC  & 2.85 ($\pm$ 0.09) & 2.50 ($\pm$ 0.10) \\
    AGAIN-VC & 2.11 ($\pm$ 0.08)  & 2.16 ($\pm$ 0.10) \\
    \hline
    \end{tabular}
    \label{tab:t1}
\end{table}

\begin{table}[!th]
	\centering
	\caption{Objective evaluation results of test accuracy (ACC), phoneme error rate (PER) and real time factor (RTF) between different models. The RTF was calculated under a single NVIDIA GeForce RTX 3090 Ti GPU.\\ } 
    \begin{tabular}{c|c|c|c}
    \hline
    Method & ACC $\uparrow$ & PER $\downarrow$ & RTF $\downarrow$  \\
    \hline
    Ground Truth      & 100\% & 2.9\% & -- \\
    StyleTTS-VC   & \textbf{91.7\%}  & {6.17\%} & 0.0128\\
    YourTTS & 49.4 \% & \textbf{5.58\%}   & 0.0369 \\
    VQMIVC & 36.0\% & 26.0\% & \textbf{0.0115}\\
    AGAIN-VC & 70.0\% & 24.6\% & 0.0143\\

    \hline
    \end{tabular}
    \label{tab:t2}
\end{table}

\subsection{Evaluations}
\label{section3.3}
We conducted subjective evaluations with two metrics: the mean opinion score of naturalness (MOS-N) which measures the naturalness of converted speech and the mean opinion score of similarity (MOS-P) which evaluates the similarity between converted and reference speech. We recruited native English speakers located in the U.S. to participate in our evaluations using an online survey through Amazon Mechanical Turk. We compared our model with two recent baseline models, AGAIN-VC \cite{chen2021again} and VQMIVC \cite{wang2021vqmivc}, and one state-of-the-art model, YourTTS\cite{casanova2022yourtts}, for any-to-any voice conversion. All baseline models were trained with official implementation \footnote{\url{https://github.com/KimythAnly/AGAIN-VC}}\footnote{
\url{https://github.com/Wendison/VQMIVC}}\footnote{\url{https://github.com/Edresson/YourTTS}} using the same train and test speaker split. For a fair comparison, the mel-spectrograms converted from all models were synthesized with HifiGAN \cite{kong2020hifi} and downsampled to 16 kHz in our evaluations.  

In every experiment, we randomly selected 40 sets of samples. When evaluating each set, we randomly permuted the order of the models and instructed the subjects to rate them without revealing the model labels. For each set, we required that there were at least five different speakers reading the same sentence, in which one was used as the ground truth and the rest four were used as the source input for our model and the three baseline models. This ensures that different samples have different lengths so that raters do not find out which one is the ground truth. The method is similar to multiple stimuli with hidden reference and anchor (MUSHRA), enabling the subjects to compare the subtle difference among models. We used the subjective rating of the ground truth as an attention check: all ratings from a subject were dropped from our analyses if the MOS of the ground truth was not ranked the highest among all models. Each set was rated by 10 raters after disqualified raters were dropped.

In addition to subjective evaluations, we also performed objective evaluations using speaker classification and phoneme error rate (PER) from an ASR model to evaluate the speaker similarity and speech intelligibility \cite{Li2021StarGANv2VCAD}. The speaker classification model consists of a ResNet-18 network that takes % 80-dimensional log filter banks 
a mel-spectrogram to predict the speaker label. The model was trained on the test speakers and we report the classification accuracy (ACC) of the trained models on samples generated with different models. We converted speech waveforms to text using an ASR model from ESPNet \cite{watanabe2018espnet} and converted the text to phoneme sequences to calculate PER.

\subsection{Ablation Study}
\label{section3.4}
To demonstrate that our approach to addressing problems in TTS-based methods is effective, we conducted an ablation study with both subjective and objective evaluations described in section \ref{section3.3}. We ablated $\mathcal{L}_{MI}$ and $\mathcal{L}_{cycle}$ when training the encoder and the decoder, respectively. In addition, to show that the latent loss introduced in Zhang et. al. \cite{zhang2021transfer} hurts the performance, we have added the loss for the encoder training defined as
\begin{equation} \label{eq7}
\mathcal{L}_{latent} = \mathbb{E}_{\bm{x}, \bm{t}}\left[{\norm{{\parn{\bm{d}_\text{align} \cdot \bm{h}_\text{text}}- { E(\bm{x})}}}_1}\right],
\end{equation}
in which the latent representation produced by $E$ is forced to be the same as that generated through text encoder and phoneme alignment. We refer to this case as $+\mathcal{L}_{latent}$ in Table \ref{tab:t2}. To demonstrate that the data augmentation for the encoder loss is effective, we also trained a model without the data augmentation. That is, we set $\hat{\bm{x}} = \bm{x}$ and only use ground truth from the training set as $\bm{x}$ in equation \ref{eq:4}.

% \begin{table}[!t]
%     \scriptsize
% 	\centering
% 	\caption{Comparison of MOS with 95\% confidence intervals (CI), test accuracy (ACC) and PER between different models. } 
%     \begin{tabular}{c|c|c|c|c}
%     \hline
%     Method & MOS-N & MOS-P & ACC & PER \\
%     \hline
%     Ground Truth      & 4.68 ($\pm$ 0.05)  & 4.58 ($\pm$ 0.07) & 100 & 2.9\\
%     StyleTTS-VC   & \textbf{3.75} ($\pm$ \textbf{0.09})  &    \textbf{3.66} ($\pm$ \textbf{0.10}) & \textbf{91.7} & \textbf{11.5}\\
%     VQMIVC  & 2.71 ($\pm$ 0.09) & 2.50 ($\pm$ 0.10) & 36.0 & 26.0\\
%     AGAIN-VC & 1.96 ($\pm$ 0.08)  & 2.16 ($\pm$ 0.10) & 70.0 & 24.6\\
%     \hline
%     \end{tabular}
%     \label{tab:t1}
% \end{table}

\section{Results}

\begin{table}[!t]
	\centering
	\caption{Subjective evaluation results of mean opinion scores (MOS) with 95\% confidence intervals (CI) between different training objectives. \\ }
	
    \begin{tabular}{c|c|c}
    \hline
    Method & MOS-N & MOS-P \\
    \hline
    Proposed    & \textbf{3.85} ($\pm$ \textbf{0.09}) & \textbf{3.67} ($\pm$ \textbf{0.11})\\
    w/o augmentation   & 3.78 ($\pm$ 0.09) & 3.67 ($\pm$ 0.12)\\
    - $\mathcal{L}_{MI}$   & 3.74 ($\pm$ 0.10) & 3.63 ($\pm$ 0.12)\\
    - $\mathcal{L}_{cycle}$  & 3.70 ($\pm$ 0.10) & 3.62 ($\pm$ 0.11)\\
    + $\mathcal{L}_{latent}$   & 3.60 ($\pm$ 0.10) & 3.58 ($\pm$ 0.11)\\
    
    \hline
    \end{tabular}
    \label{tab:t3}
\end{table}
Table \ref{tab:t1} and \ref{tab:t2} show the results of comparison between different models and the ground truth. Our model significantly outperforms the other baseline models in both naturalness and similarity in the subjective evaluation experiment. Our model also scored higher in testing accuracy and PER than other baseline models except for YourTTS in PER. However, we do note that the difference is small as shown in Table \ref{tab:t2}. In addition, since our model is not Flow-based, we do not need to compute the Jacobian and matrix inversion required by Flow-based YourTTS. This makes our model significantly faster than YourTTS as indicated by RTF in Table \ref{tab:t2}. 
The ablation study results in Table \ref{tab:t3} and \ref{tab:t4} show that removing $\mathcal{L}_{MI}$ or $\mathcal{L}_{cycle}$ decreases both naturalness and similarity of the synthesized speech. The baseline model with full objectives also outperforms models trained without $\mathcal{L}_{MI}$ or $\mathcal{L}_{cycle}$ in classification accuracy and PER. Training without data augmentation also decreases the rated naturalness and objective metrics. 

It is worth noting that the rated naturalness and similarity drop significantly when we add the proposed $\mathcal{L}_{latent}$ in \cite{zhang2021transfer}. We hypothesize that by enforcing $\bm{d}_\text{align} \cdot \bm{h}_\text{text} = E(\bm{x})$ through $\mathcal{L}_{latent}$, we essentially obtain an ASR model because $E$ is trained to produce an aligned version of $\bm{h}_\text{text}$ which consists of merely phoneme token embeddings. We show an example of inverted alignment using $E(\bm{x})$ to illustrate our hypothesis. As shown in Figure \ref{fig:fig2}, the inverted alignment $E(\bm{x}) \cdot \bm{h}_{\text{text}}^{-1}$ successfully reconstructs the monotonic alignment $d_\text{align}$ with some noise when $E$ is trained with $\mathcal{L}_{latent}$. This shows that $E(\bm{x})$ consists roughly of discretized phoneme representations as $d_\text{align}$ can be recovered through a pseudoinverse of $\bm{h}_\text{text}$. On the other hand, the encoder trained without $\mathcal{L}_{text}$ fails to recover $d_\text{align}$ with $\bm{h}_\text{text}^{-1}$, indicating that $E$ learns a different representation that the decoder can use to reconstruct the natural speech produced by monotonic alignment and text representation. Training without $\mathcal{L}_{latent}$ avoids the problems associated with ASR models such as incorrectly recognized phonemes that can make speech unclear or produce incorrect phonetic content. 

% \begin{table}[!t]
%     \scriptsize
% 	\centering
% 	\caption{Reported MOS with 95\% confidence intervals (CI), test accuracy (ACC) and PER between different training objectives. }
	
%     \begin{tabular}{c|c|c|c|c}
%     \hline
%     Method & MOS-N & MOS-P & ACC & PER \\
%     \hline
%     Baseline    & \textbf{3.71} ($\pm$ \textbf{0.09}) & \textbf{3.67} ($\pm$ \textbf{0.11}) & \textbf{91.7} & \textbf{11.5}\\
%     - $\mathcal{L}_{asr}$   & 3.60 ($\pm$ 0.10) & 3.63 ($\pm$ 0.12) & 91.4  & 14.9\\
%     - $\mathcal{L}_{cycle}$  & 3.56 ($\pm$ 0.10) & 3.62 ($\pm$ 0.11) & 90.0  & 17.1\\
%     + $\mathcal{L}_{latent}$   & 3.46 ($\pm$ 0.10) & 3.58 ($\pm$ 0.11) & 91.2  & 20.6\\
%     \hline
%     \end{tabular}
%     \label{tab:t2}
% \end{table}

\begin{table}[!t]
	\centering
	\caption{Objective evaluation results of speaker-classification test accuracy (ACC) and phoneme error rate (PER) between different training objectives. \\ }
	
    \begin{tabular}{c|c|c}
    \hline
    Method & ACC $\uparrow$ & PER $\downarrow$ \\
    \hline
    Baseline & \textbf{91.7\%} & \textbf{10.4\%}\\
    w/o augmentation   & 90.5\%  & {11.5}\%\\
    - $\mathcal{L}_{MI}$   & 91.4\%  & 14.9\%\\
    - $\mathcal{L}_{cycle}$  & 90.0\%  & 17.1\%\\
    + $\mathcal{L}_{latent}$   & 91.2\%  & 20.6\% \\
    \hline
    \end{tabular}
    \label{tab:t4}
\end{table}

\section{Conclusions}
We propose a framework using a style-based TTS model for one-shot voice conversion with novel cycle consistency and phoneme MI maximization objectives in place of the latent reconstruction objective. The framework employs a novel data augmentation scheme that fully explores the input and output space of pre-trained TTS decoders. The proposed model achieves state-of-the-art performance in similarity and naturalness with both subjective and objective evaluations where our models scored significantly higher in various metrics, including MOS, ACC, and PER than previous models. We also demonstrate that using the latent reconstruction loss $\mathcal{L}_{latent}$ proposed in \cite{zhang2021transfer} worsens speech similarity and naturalness and we illustrate a potential explanation for this effect. Moreover, unlike other one-shot voice conversion systems such as \cite{wang2021vqmivc} and \cite{zhang2021transfer}, our framework is completely convolutional and can therefore perform real-time inference with a faster-than-real-time vocoder. However, we acknowledge that our framework requires text labels during training, which can be prohibitive to train on unannotated large-scale speech corpora. Future work includes removing the need for text labels through semi-supervised or self-supervised learning. We would also like to improve speaker similarity by learning a better speaker representation that can reproduce the accent of unseen speakers. 

\section{Acknowledgments}
This work was funded by the national institute of health (NIHNIDCD) and a grant from Marie-Josee and Henry R. Kravis. 

% -------------------------------------------------------------------------
\bibliographystyle{IEEEbib}
\bibliography{mybib}

\end{document}